**Title:** Bioimpedance a Diagnostic Tool for Tobacco Induced Oral Lesions – a Mixed Model cross-sectional study

**Running title:** Bioimpedance a Diagnostic Tool – a cross-sectional study

**Authors**

1. Dr. Vaibhav Gupta

Young Scientist

National Institute of Pathology - ICMR

Delhi-110029, India

Mail id: drvaibhavphd@gmail.com

ORCID ID: 0000-0001-5272-0438

2. Dr. PoonamGoel

Women Scientist

National Institute of Pathology - ICMR

Delhi-110029, India

Mail id: dr.poonam1187@gmail.com

ORCID ID: 0000-0002-9767-0283

3. Dr. UshaAgrawal

Ex- Director

National Institute of Pathology - ICMR

Delhi-110029, India

Mail id: uburra@gmail.com

4. Dr. Neena Chaudhary

Professor and HOD

Department of ENT

VMMC and Safdarjung Hospital

Delhi-110029, India

Mail id: drneenachaudhary@hotmail.com

5. Dr.Garima Jain

Scientist D,

National Institute of Pathology - ICMR

Delhi-110029, India

Mail id: lhmc.garima@gmail.com

6. Dr. Deepak Gupta

Associate Professor

Department of ENT; VMMC and Safdarjung Hospital

Delhi-110029, India

Mail id: guptadeepak74@yahoo.co.in

**Corresponding Author:**

Dr. Vaibhav Gupta

Young Scientist

National Institute of Pathology - ICMR

Delhi-110029, India

Mail id: drvaibhavphd@gmail.com

**Declarations Section:**

Conflict of interest: Nil

Source of Funding: This research received a grant from the Department of Health Research, Delhi, India, under the HRD scheme (R.12014/34/2021-HR/E-Office: 8114747).

Ethical Approval: Ethical clearance was obtained from the institutional ethical committee of the National Institute of Pathology (NIP-IEC/26-05-2022 /01/01R2) and VMMC &Safdurjung Hospital (IEC/VMMC/SJH/Project/2022-03/CC-241).

Informed Consent: obtained from all the subjects

Abstract word count: 138

Manuscript word count: 3245

Number of references: 16

Number of figures/tables: 8

Figures: 5

Tables: 3

**Statement of Clinical Relevance:** Successful management of cancer depends on proper screening and treatment methods. Bioimpedance appears to be a promising tool for oral cancer screening at community level due to its non-invasiveness, reliability, immediate results, low cost and portability of whole system.

# Bioimpedance a Diagnostic Tool for Tobacco Induced Oral Lesions – a Mixed Model cross-sectional study

**Abstract:**

**Introduction:** Electrical impedance spectroscopy (EIS) has recently developed as a novel diagnostic device for screening and evaluating cervical dysplasia, prostate cancer, breast cancer and basal cell carcinoma. The current study aimed to validate and evaluate bioimpedance as a diagnostic tool for tobacco-induced oral lesions.

**Methodology:** The study comprised 50 OSCC and OPMD tissue specimens for in-vitro study and 320 subjects for in vivo study. Bioimpedance device prepared and calibrated. EIS measurements were done for the habit and control groups and were compared.

**Results:** The impedance value in the control group was significantly higher compared to the OPMD and OSCC groups. Diagnosis based on BIS measurements has a sensitivity of 95.9% and a specificity of 86.7%.

**Conclusion:** Bioimpedance device can help in decision-making for differentiating OPMD and OSCC cases and their management, especially in primary healthcare settings.

**Introduction:**

Globally, oral cancer is in the sixth position in prevalence among tumors and showed the highest prevalence among head and neck cancer.[1] Furthermore, 33% of the global prevalence is in the Indian subcontinent.[2] Despite the latest treatment modalities, the 5-year survival rate has not improved, primarily due to late diagnosis. Oral squamous cell carcinoma (OSCC) could be preceded by oral potentially malignant disorders (OPMDs), a collection of clinically suspicious lesions.[1] Most of them are diagnosed as white lesions (leukoplakia) or, infrequently, as red lesions (erythroplakia) on the oral mucosa. Worldwide, the incidence of OPMDs ranges from 0.6 to 30.2/1,000 people, and prevalence fluctuates between 1–5% in different regions.[3,4] Studies have shown that OPMD can transform into OSCC at six months.[1] Timely finding and referral is a keystone to increasing survival and decreasing investigative delay.[2] Presently, visual and tactile examination of oral lesions, followed by tissue biopsy, is still the gold standard for diagnosing OPMDs and OSCC.[3]

Many novel methods are available for timely diagnosis of OSCC, each with its qualities and shortcomings, but so far unsuccessful in its practical implication in the community system. Moreover, specific procedures like vital staining (Toluidine blue) is associated with subjective bias.[1,4] The chief drawback of biopsy and histopathological investigation for evaluating OPMDs is that many lesions express no epithelial dysplasia, so several patients may get a biopsy unreasonably, leading to unfavourable physical and psychological consequences for patients and their families. It also upsurges the cost of treating OPMDs.[3] Recently, increasing awareness has been seen in developing electrical and optical devices for OPMD evaluation. Imaging procedures like optical coherence tomography and narrow band imaging that can distinguish alteration in the cellular structure of oral epithelium have been used to evaluate OPMD.[1,3]

BIS (Bioimpedance spectroscopy) is precisely analogous to Electrical impedance spectroscopy (EIS); however, in BIS, the method is solely for biological tissues.[5] BIS has been developed with interdisciplinary research and is non-invasive compared to other screening procedures. BIS has recently developed as a novel diagnostic device for screening and evaluating cervical dysplasia,[6] prostate cancer,[7] basal cell carcinoma[8] and breast cancer[9,10]. Bioimpedance can be defined as the organic tissue's reaction to resist an external electric current.[11] The frequency of electric current alters the electric properties of cells defined as alpha, beta and gamma dispersion. The ionic atmosphere near the cells affects the alpha dispersion at low frequencies between 10 Hz to 10 kHz. The beta dispersion displays structure relaxation at a frequency between 10 kHz to 10 MHz. The gamma dispersion attained at higher frequencies is related to water molecules.[4,11]

While recording bioimpedance, the current frequency is set at 50 kHz, as a nominal current of this frequency cannot excite tissues such as nerve cells and cardiac muscles. Practically, the patient cannot feel the current magnitude of around 800 μA.[12] Many researchers advised its application in screening breast and cervical cancer because of its properties, like its economical, non-invasive, portable, and immediate and reliable results.[6,9]

To our knowledge, no research has explored and validated the bioimpedance value associated with tobacco-induced oral lesions. Hence, the study aimed to validate and evaluate bioimpedance as a diagnostic tool for tobacco-induced oral potentially malignant disorders and oral squamous cell carcinoma.

**Methodology**

A cross-sectional study (in-vitro and in-vivo)was conducted at the National Institute of Pathology, New Delhi and the Dept. of ENT, VMMC &Safdurjung Hospital, New Delhi, in compliance with the Helsinki Declaration. Ethical clearance was obtained from the institutional ethical committee of the National Institute of Pathology *(NIP-IEC/26-05-2022 /01/01R2)* and VMMC &Safdurjung Hospital *(IEC/VMMC/SJH/Project/2022-03/CC-241)*.

**Study population:** The study comprised 50 OSCC and OPMD tissue specimens for in-Vitro study and 320 subjects in total for in vivo study (estimated after setting the level of significance at 95%, prevalence of OPMD in Indian population as 7.3% as per National oral health survey and Fluoride mapping[13] and level of precision as 6%). The subjects were recruited using random sampling from the outpatient department of ENT, Safdurjung Hospital. They were divided into four groups: subjects diagnosed with OSCC, subjects diagnosed with OPMD, subjects with tobacco habit but no lesion and a control group comprised of subjects with no habit of tobacco consumption (relatives of the subjects with tobacco-induced lesion). Medical & dental history with intraoral clinical examination was recorded, and written informed consent was obtained. Subjects with a history of tobacco usage were included in study groups. Furthermore, subjects with a history of systemic diseases, patients with pacemakers and pregnant and lactating females were excluded.

**Data Collection:** was conducted in three phases:

**Phase I**:

**Bioimpedance measurement:**

The prepared device consisted of a plastic probe connected to the evaluation board with a high-precision impedance converter system (Analog Devices, Wilmington) (figure 1a) and Evaluation Software installed on a laptop to download the data. The device was calibrated, and all the parameters were calculated using Evaluation Software. The probe tip has electrodes to attain a definite current stream into the tissue. A probe was positioned softly on the tobacco-induced lesion (study groups) or healthy oral mucosa (control group) to record the impedance, as shown in Figure 1b. In this procedure, the current is passed at a frequency of 50kHz with an output excitation voltage range of 0.2V to measure four electrical properties: Impedance (Z), Real part of impedance (R), Imaginary part of impedance (X); and Phase angle ($\theta$).

Before using the device on patients, the principal investigator was trained. As part of an in-vitro study, bioimpedance reading was recorded for 50 biopsied OSCC and OPMD tissue specimens with sizes ranging from 0.5 to 1 cm. Biopsied tissue specimens were transferred to the 0.9% saline solution immediately after excision to maintain the isogenic environment. Three distinct measurements of each site were made in series to assess the reliability of measurements, and the average of the three readings was used for analysis. Data was recorded and downloaded on the laptop for each patient in real time for further analysis.

**Phase II:**

Study subjects were allocated to the groups as per the presence/ absence of habit and presence/ absence of lesion. Furthermore, bioimpedance value was assessed for all the groups and compared. Three distinct measurements of each site were made in series at the same point to assess the reliability of measurements, and the average of the three readings was used for analysis.

**Phase III:**

Tobacco cessation counselling was done for subjects with a tobacco habit; subjects with no habits were motivated not to start the habit.

**Statistical analysis:** Bioimpedance values assessed for all four groups were entered in Microsoft *Excel. In-vitro* data were analyzed using an unpaired t-test, *in-vivo* data wereanalyzed for intergroup comparison using ANOVA with post hoc Tukey test; for risk factors and demographic variables, the chi-square test was performed using SPSS software v.21, and also the sensitivity and specificity of bioimpedance was assessed comparing with biopsy (Gold standard), and the ROC curve was plotted. The statistically significant difference was considered if $p \leq 0.05$.

**Results:**

***In-vitro Analysis:*** The difference was insignificant when comparing histopathological diagnostic reports with impedance diagnostic results for biopsied specimens using the unpaired t-test. The average impedance score of 27 biopsied specimens was $349\pm58.21$, which was histopathologically reported as OSCC, and for the remaining 23 specimens, it was $583\pm17.29$, which was histopathologically reported as OPMD. Once the tool was established, in-vivo data collection was done and analyzed.

***Demographic data:*** In the study conducted male, to female ratio was 2:1, and the statistical difference between groups was not significant (p=0.79). Subjects recruited were aged between 16 to 82 years. For OPMD and OSCC groups, most of the subjects were in the fifth decade of life with mean ages of $48.09\pm12.06$years and $49.50\pm16.82$ years, respectively (Table2), and the difference between groups was statistically insignificant (p=0.89).

***Risk factors:*** Alcohol consumption was quite significant among all the habit groups. Whereas other risk factors distribution among the four groups was not significantly different (p>0.05), as shown in Table 2. Due to pain in the lesion, 18 subjects could not eat properly. Surprisingly 4% of subjects were addicted to narcotic drugs, which were referred to the concerned department for necessary management.

Clinically most of the lesions were present on the tongue (27%), followed by buccal mucosa (23.5%), alveolar-buccal sulcus (21%) and labial mucosa (17%). Among OSCC subjects, 28% had proliferative growth, 35% had ulcerative lesions, and 37% had ulceroproliferative patterns. Based on TNM staging, 18% were in Stage I, 10% in Stage II, 30% in Stage III, 37% in Stage IVa, 3% in Stage IVb and 2% in Stage IVc. Among OPMD subjects, around 79.5% had white lesions suspecting leukoplakia, 12% had red lesions (erythroplakia), and the remaining 8.5% showed pseudomembranous growth. Histopathologically 29% of OPMD lesions showed severe dysplasia, and 41% showed mild dysplasia. Moreover, in the OSCC group, 49% of lesions were well differentiated, 37% were moderately differentiated, and 14% were poorly differentiated.

**BIS measurement:**

*Impedance*: In the control group, the impedance value ranged from 921.95 to 1118.03 with a mean of 1054.48±47.69 whereas, among subjects with the habit but no lesion, the value ranged from 707.11 to 921.95 with a mean value of 896.83±56.12. In the lesion group, impedance was low compared to the control group; the value ranged from 318.25 to 621.78 in the OPMD group and from 101.22 to 515.38 in the OSCC group, with an average of 472.26±86.65 and 335.45±92.70 respectively, (figure 2). The difference among the four groups was found to be statistically significant (table 3), and the post hoc Tukey test showed significant differences between different combinations of the two groups ($p<0.001$). Table 3 shows the mean value along with the standard deviation for the variables of the bioimpedance device. The readings in the no habit no lesion and habit no lesion groups are slightly different but significantly different from the OPMD and OSCC group readings.

*Phase angle:* In the control group, the value ranged from -139.40 to -81.67 with a mean of -107.74±17.72 whereas, among subjects with the habit but no lesion, the value ranged from -139.39 to -18.44 with a mean value of -90.8±32.58. In the lesion group, the phase angle was high compared to the control group; the values ranged from -90.41 to -4.40 in the OPMD group and from -69.27 to 8.13 in the OSCC group with an average of -24.15±17.89 and -15.04±14.52 respectively (figure 2). The difference among the four groups was found to be statistically significant (table 3) ($p<0.001$).

*The real part of Impedance:* In the control group, the value ranged from 319.84 to 670.82 with a mean of 548.53±143.30 whereas, among subjects with the habit but no lesion, the value ranged from 249.99 to 553.17 with a mean value of 519.25±69.83. In the lesion group phase angle was smaller compared to the control group; the values ranged from 101.81 to 325.96 in the OPMD group and from 32.01 to 230.09 in the OSCC group, with an average of 156.17+33.58 and 143.03+57.96 respectively (figure 2). The difference among the four groups was found to be statistically significant (table 3) (p<0.001).

*The imaginary part of Impedance:* In the control group, the value ranged from -1000.03 to -824.62 with a mean of -890.26+62.46 whereas, among subjects with the habit but no lesion, the value ranged from -824.65 to -632.46 with a mean of the value of -780.89+70.46. In the lesion group, the phase angle was more compared to the control group; the values ranged from -977.87 to 254.60 in the OPMD group and from -488.93 to 254.60 in the OSCC group with an average of -432.35+135.07 and -288.77+123.20 respectively (figure 2). The difference among the four groups was found to be statistically significant (table 3) (p<0.001).

**Assessment and validation of BIS**

The ROC curve was prepared using BIS statistics to evaluate the accuracy of the BIS measurement to recognize tobacco-induced oral lesions. Figure 3 shows how accurately BIS can distinguish the presence or absence of lesions (OSCC & OPMD)in comparison to biopsy with the area under the curve (AUC) of 0.913 (95% confidence intervals, 0.810 - 1.0; p<0.001). The highest point in the ROC curve corresponds to a sensitivity of 95.9% and a specificity of 86.7%.

Figure 3b demonstrates the performance of BIS in differentiating high-risk lesions (poorly to moderately differentiated OSCC&moderate to severe dysplasia) from low-risk lesions (well differentiated OSCC & mild dysplasia) with an area under the curve (AUC) of 0.907 (95% confidence intervals, 0.829–0.985; p<0.001). The highest point in the ROC curve equates to a sensitivity of 93.3% and a specificity of 88%.

**Discussion**

The first study on bioimpedance in oncology was conducted in 1926.[4] In literature, bioimpedance has been considered a screening device for different cancers all over the body, but there is limited research on OSCC. Various frequencies ranging between 20 Hz–5 MHz have been assessed to measure impedance for diagnosing oral cancer. It was concluded that at a frequency of 20 Hz and 50 kHz, OSCC could be distinguished from healthy tissue.[1,11,14,15] As per a detailed literature search, no research has explored and validated the bioimpedance value associated with tobacco-induced oral lesions.[1,11,14,15,16] Therefore, the study was conducted to validate and evaluate bioimpedance as a diagnostic tool for tobacco-induced oral lesions.

Sarode et al. prepared a bioimpedance device using an ice cream stick as probe 1, whereas Murdoch et al. used a bioimpedance tool comprised of a handheld unit and a base station for saving data to the system.[3] Tatullo et al. used a probe within the "head" containing four micro-electrodes.[16] In the study, we prepared a BIS device consisting of a plastic probe connected to the evaluation board with a high-precision impedance converter system (figure 1) and Evaluation Software installed on a laptop for downloading the data.

Before using the device on patients, the device was calibrated, and the principal investigator was trained. Bioimpedance measurement was initially done for 50 biopsied tissue specimens, and no statistically significant difference in the diagnosis was obtained for the BIS compared to the gold standard.

No difference in the BIS measurement was seen based on gender, religion and marital status among control and habit groups. Ethanol consumption showed an impact on BIS reading which needs to be further explored. Other risk factors showed non-significant distribution among groups. Hence their effect on BIS was not seen.

No significant difference was seen based on anatomical sites in all four groups. Still, labial and buccal mucosa showed higher impedance, whereas the hard palate had comparatively lower impedance. However, Murdoch et al. displayed variance among the nine anatomical sites at frequencies between 1.2–39 kHz.[3] Significantly negative correlation was seen between TNM staging and impedance measurement.

The impedance value in the control group ranged from 921.95 to 1118.03, and in the habit but no lesion group, the value ranged from 707.11 to 921.95, which is significantly higher compared to the OPMD group, the value ranged from 318.25 to 621.78 and OSCC group from 101.22 to 515.38. The results obtained are in accordance with the findings observed by Sarode et al.,[1] Ching et al.[14] and Sun et al.[15] In the study, even the impedance measurement of the OSCC group was significantly lower compared to the OPMD group. Furthermore, the BIS measurement of the no habit no lesion group was significantly higher than the habit no lesion group. In both OPMD and OSCC groups, high-risk lesions (poorly to moderately differentiated OSCC & moderate to severe dysplasia) had significantly lower impedance than low-risk lesions (well differentiated OSCC & mild dysplasia) with a sensitivity of 93.3%. Hence, patients with low-risk lesions can be appropriately managed based on impedance measurement, and transformation to high-risk can be prevented. Even the overall 5-year survival of oral cancer patients can be increased via timely diagnosis using a BIS device. It shows tobacco consumption can reduce the impedance by pathological changes in the tissue which are clinically not appreciable and can help the clinician in counselling patients for tobacco cessation.

In the control group, the phase angle value ranged from -139.40 to -81.67 and habit but no lesion group, the value ranged from -139.39 to -18.44, which is significantly lower compared to OPMD and OSCC group, the value ranged from -90.41 to -4.40 and from -69.27 to 8.13 respectively. Sarode et al.,[1] Ching et al.,[14] and Sun et al.[15] reported similar findings.

The real part of the Impedance value in the control group value ranged from 319.84 to 670.82, and in habit but no lesion, ranged from 249.99 to 553.17 is significantly higher compared to OPMD and OSCC groups. Furthermore, the Imaginary part of the Impedance value in the control group had a mean of -890.26±62.46, and the habit but no lesion group had a mean of -780.89±70.46, which is significantly lower compared to the OPMD and OSCC groups.

Accuracy of diagnostic results of BIS in comparison to histopathology is obtained from the ROC plot with an AUC of 0.913 ($p<0.001$) with a sensitivity of 95.9% and specificity of 86.7% (figure 3a). It concludes bioimpedance device is valid enough as a diagnosis based on BIS measurements had a high percentage of true positive results, and very few cases were false negative. Hence, it can be used in public health settings with limited resources, especially money and human resources. Biopsy can be avoided in public settings, saving resources, especially time and money. It can be easily carried anywhere due to its lightweight, making it a highly recommendable device for community settings. Even with little training, paramedics can easily use the device in suspected cases. Immediate results can omit the waiting time, and the test can be repeated in case of any uncertainty. Moreover, non-invasive testing can reduce apprehension among patients and increase compliance. Even patients with tobacco habits but no lesions can be tested for change in impedance measurement and used for tobacco cessation counselling as a preventive tool.

Risk factors were not controlled while measuring BIS for tobacco-induced oral cancer can be a study limitation. Further research is needed to explore the role of various risk factors concerning tobacco on BIS measurement.

**Conclusion:**

Bioimpedance, a novel diagnostic tool, can be used for management by timely diagnosing patients with low-risk lesions, improving their five-year survival, and decreasing the five-year mortality rate. Moreover, it can also be used for preventive management by tobacco cessation counselling of patients with habit and no lesions.

*Author Contribution:*
- V.G.: Conception and design of the work, Device preparation, Patient screening and collection of data, Data acquisition, analysis, or interpretation, Preparation of the manuscript
- P.G.: Patient screening and collection of data, histopathology reporting and data acquisition and analysis
- N.C.: Revising manuscript critically for important intellectual content, Clinical patient management
- G.J.: Histopathology reporting, revising manuscript critically for important intellectual content.
- D.G.: Revising manuscript critically for important intellectual content, Clinical patient management
- U.A.: Histopathology reporting, revising manuscript critically for important intellectual content, final approval of the version to be published.
- All the authors reviewed the manuscript

*Ethics Approval:* Ethical clearance was obtained from the institutional ethical committee of the National Institute of Pathology *(NIP-IEC/26-05-2022 /01/01R2)* and VMMC &Safdurjung Hospital *(IEC/VMMC/SJH/Project/2022-03/CC-241)*.

*Consent to Participate:* obtained from all the subjects

*Conflict of interest:* Nil

*Source of Funding:* This research received a grant from the Department of Health Research, Delhi, India, under the HRD scheme (R.12014/34/2021-HR/E-Office: 8114747).

**Table 1: Patient distribution as per various demographic factors among four groups**

| | | Group | | | | Chi-square value | Sig.* |
|---|---|---|---|---|---|---|---|
| | | Habit no lesion | No habit, no lesion | OPMD | OSCC | | |
| Gender | Female | 28 | 25 | 31 | 27 | 1.035 | 0.79 |
| | Male | 52 | 55 | 49 | 53 | | |
| Marital Status | Married | 62 | 55 | 59 | 63 | 4.369 | 0.89 |
| | Single | 16 | 23 | 18 | 14 | | |
| | Widow | 1 | 1 | 2 | 1 | | |
| | Divorced | 1 | 1 | 1 | 2 | | |
| Religion | Hindu | 53 | 50 | 55 | 48 | 4.014 | 0.67 |
| | Islam | 22 | 28 | 23 | 29 | | |
| | Sikh | 5 | 2 | 2 | 3 | | |
| Age** (years) | | 39±7.62 | 41.46±13.26 | 48.09±12.06 | 49.50±16.82 | | 0.21 |

*difference is statistically significant if value<0.05
**unpaired t-test

**Table 2: Risk factors distribution for oral carcinoma among four groups**

| | | Group | | | | Sig.* |
|---|---|---|---|---|---|---|
| | | Habit no lesion | No habit, no lesion | OPMD | OSCC | |
| Tobacco use | No | 0 | 80 | 0 | 0 | <0.001 |
| | Yes | 80 | 0 | 80 | 80 | |
| Alcohol use | No | 25 | 51 | 13 | 17 | <0.001 |
| | Yes | 55 | 29 | 67 | 63 | |
| Narcotic drugs intake | No | 76 | 79 | 75 | 77 | 0.42 |
| | Yes | 4 | 1 | 5 | 3 | |
| Diet | Mixed | 46 | 49 | 44 | 52 | 0.59 |
| | Vegetarian | 34 | 31 | 36 | 28 | |
| Vitamin-rich diet | Yes | 77 | 75 | 72 | 70 | 0.18 |
| | No | 3 | 5 | 8 | 10 | |
| Spicy Food Intake | No | 40 | 46 | 44 | 38 | 0.57 |
| | Yes | 40 | 34 | 36 | 42 | |
| Hot food intake | No | 75 | 78 | 76 | 69 | 0.17 |
| | Yes | 5 | 1 | 4 | 11 | |
| Sharp tooth | Yes | 1 | 3 | 2 | 6 | 0.18 |
| | No | 79 | 77 | 78 | 74 | |

*difference is statistically significant if value<0.05 on the chi-square test

**Table 3: ANOVA showing the comparison between all four groups for all the BIS variables**

|  |  | Mean | Std Dev. | Sig.* |
|---|---|---|---|---|
| Impedance | No Habit, No Lesion | 1054.48 | 47.69 | <0.001 |
|  | Habit No Lesion | 896.83 | 56.12 |  |
|  | OPMD | 472.26 | 86.65 |  |
|  | OSCC | 335.45 | 92.70 |  |
| Phase angle | No Habit, No Lesion | -107.74 | 17.72 | <0.001 |
|  | Habit No Lesion | -90.79 | 32.58 |  |
|  | OPMD | -24.15 | 17.89 |  |
|  | OSCC | -15.04 | 14.52 |  |
| The real part of the impedance | No Habit, No Lesion | 548.53 | 143.30 | <0.001 |
|  | Habit No Lesion | 519.25 | 69.83 |  |
|  | OPMD | 156.17 | 33.58 |  |
|  | OSCC | 143.03 | 57.96 |  |
| The imaginary part of the impedance | No Habit, No Lesion | -890.26 | 62.46 | <0.001 |
|  | Habit No Lesion | -780.89 | 70.46 |  |
|  | OPMD | -432.35 | 135.07 |  |
|  | OSCC | -288.77 | 123.20 |  |

*difference is statistically significant if value<0.05

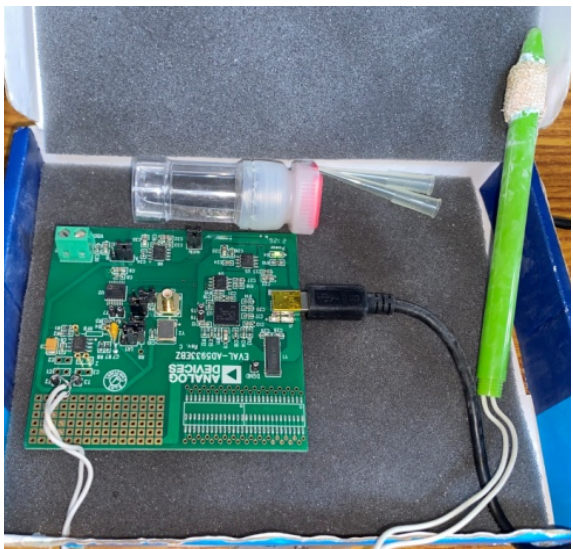

**Figure 1a: Bioimpedance device**

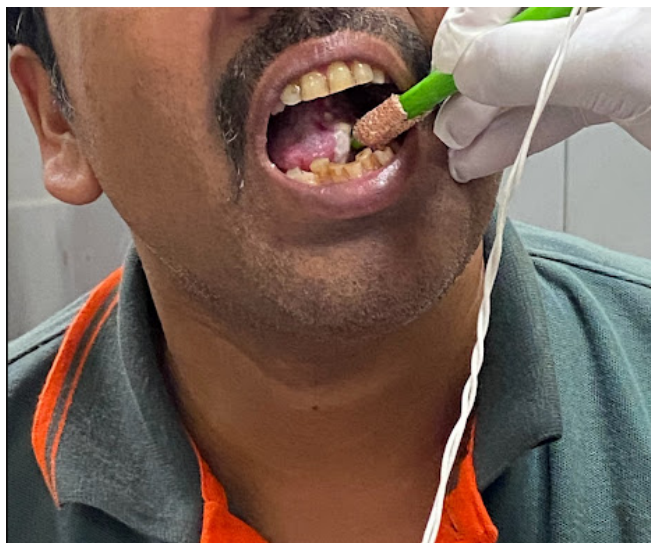

**Figure 1b: Recording measurement using a probe of bioimpedance device**

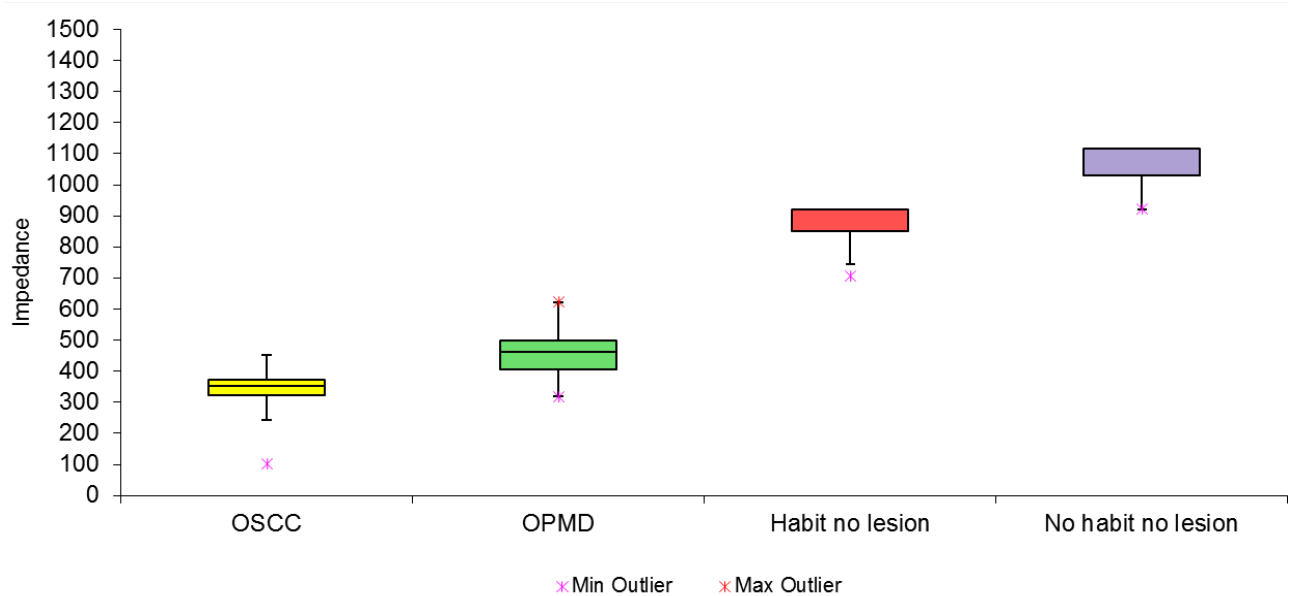

**Figure 2: Box and whisker plot shows the distribution of readings across four groups**

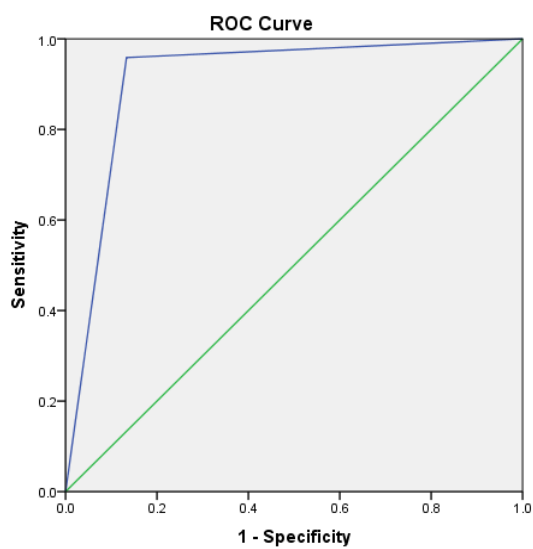 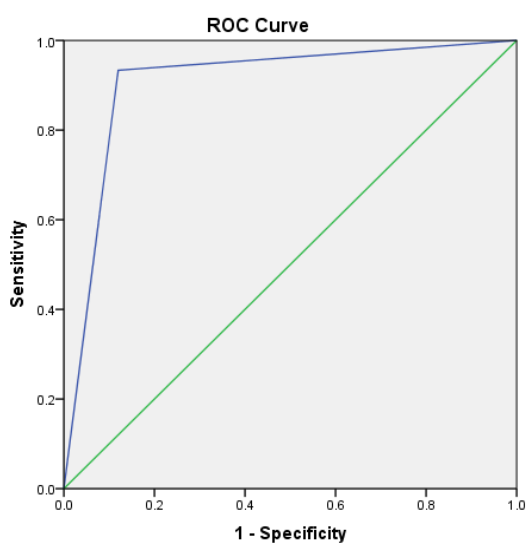

**Fig 3a: ROC curve showing the accuracy of BIS in identifying the presence of lesions in comparison to biopsy**

**Fig 3b: ROC curve showing BIS's performance in differentiating high-risk and low-risk lesions.**